\documentclass[runningheads]{llncs}

\usepackage[T1]{fontenc}
\usepackage{graphicx}

\usepackage[x11names,dvipsnames]{xcolor}
\usepackage{microtype}
\usepackage{amsmath}
\usepackage{amssymb}
\usepackage{booktabs}
\usepackage{tabularx}
\usepackage{listings}
\usepackage{algorithm}
\usepackage{algpseudocode}
\usepackage{pgfplots}
\pgfplotsset{compat=1.18}
\usetikzlibrary{positioning,arrows.meta,shapes.geometric,shapes.arrows,calc,decorations.pathreplacing,fit}
\definecolor{distbarmain}{HTML}{2C6FBB}
\definecolor{distbaralt}{HTML}{B0B7BE}
\definecolor{distbarwarn}{HTML}{C44E52}
\pgfplotsset{
  distbar/.style={
    xbar, y dir=reverse,
    width=0.92\linewidth,
    bar width=8pt,
    xmin=0,
    enlarge y limits=0.12,
    axis x line*=bottom, axis y line*=left,
    tick align=outside,
    ytick=data,
    yticklabel style={font=\small},
    xticklabel style={font=\footnotesize},
    xlabel style={font=\footnotesize},
    nodes near coords, nodes near coords align={horizontal},
    every node near coord/.append style={font=\scriptsize},
    every axis plot/.append style={fill=distbarmain, draw=distbarmain!80!black},
  },
}
\definecolor{codebg}{HTML}{F6F8FA}
\definecolor{coderule}{HTML}{D0D7DE}
\definecolor{codecomment}{HTML}{6A737D}
\definecolor{codestring}{HTML}{1B7F4B}
\lstdefinestyle{paper}{
  language=Python,
  basicstyle=\ttfamily\small,
  keywordstyle=\color{distbarmain}\bfseries,
  commentstyle=\color{codecomment},
  stringstyle=\color{codestring},
  showstringspaces=false,
  breaklines=true,
  breakatwhitespace=true,
  columns=fullflexible,
  keepspaces=true,
  backgroundcolor=\color{codebg},
  frame=single,
  rulecolor=\color{coderule},
  framerule=0.5pt,
  framesep=6pt,
  xleftmargin=9pt,
  xrightmargin=3pt,
  aboveskip=\medskipamount,
  belowskip=\medskipamount,
  tabsize=2,
}
\usepackage[hidelinks]{hyperref}

\title{``My Credentials, Please'': Detecting Filesystem Capability Laundering in MCP Servers at Scale}
\titlerunning{Detecting Filesystem Capability Laundering in MCP Servers}

\author{%
Zeqian~Huang\inst{1} \and
Mingyue~Tang\inst{2} \and
Aonan~Guan\inst{3} \and
Qiming~Sun\inst{4} \and
Pengzhou~He*\inst{5}}
\authorrunning{Z. Huang et al.}
\institute{%
Independent Researcher\\
\email{leilahuang95@gmail.com}
\and
University of Illinois Urbana-Champaign, Urbana, IL, USA\\
\email{mt55@illinois.edu}
\and
Independent Researcher\\
\email{guanaonan@gmail.com}
\and
Santa Clara University, Santa Clara, CA, USA\\
\email{qsun4@scu.edu}
\and
Auburn University at Montgomery, Montgomery, AL, USA\\
\email{phe@aum.edu}}

\begin{document}
\pagestyle{plain}

\maketitle
\thispagestyle{plain}

\begin{abstract}
The Model Context Protocol (MCP) standardizes how AI agents invoke external tools. MCP servers expose tools through a declared contract consisting of a name, description, and argument schema. In typical local deployments, client-side approval of a tool call grants the underlying server process broad host filesystem permissions rather than restricting it to the declared operation.
We study the resulting vulnerability class, \emph{filesystem capability laundering}, where attacker-steered arguments cause a tool to access sensitive host files outside its declared contract and deployment scope.
We analyze this capability gap across three layers: server-category screening (L1), per-tool contract analysis (L2), and bypassable path-validation checks (L3).
We design MCP-FSD to catalog MCP servers, infer least-privilege boundaries per tool, plant credential fixtures outside those boundaries, and evaluate exploitability and sandbox containment via paired dynamic execution.
Across 21{,}097 MCP servers from nine registries, 787 of 4{,}806 triaged servers (16.4\%) exhibit filesystem capabilities despite non-filesystem functional classifications. Dynamic three-case validation confirms 150 vulnerable instances under tested sandbox policies. Coordinated disclosure yielded 13 security advisories, including two CVEs in Anthropic's Git MCP Server.

\end{abstract}

\section{Introduction}
\label{sec:introduction}

The Model Context Protocol (MCP)~\cite{mcp-landscape} has become the standard interface through which AI agents invoke external tools, data sources, and services. An MCP server implements tools that a client discovers through \texttt{tools/list} and invokes through \texttt{tools/call}. The tool's name, description, and argument schema form its \emph{declared contract}~\cite{mcp-tools}.

MCP supports client-provided filesystem roots~\cite{mcp-roots}, and tool definitions can include output schemas and behavioral annotations~\cite{mcp-tools}. However, the MCP specification delegates per-tool filesystem enforcement entirely to individual server implementations or host environments. The security question is whether attacker-influenced arguments can steer a tool's effects beyond the scope justified by its contract and deployment context.

Path-traversal flaws are documented in the agent harness itself: Claude Code's prefix-match bypass~\cite{claude-code-pt1} and symlink-following~\cite{claude-code-pt2} let crafted arguments escape the harness's intended scope. For locally launched servers, the code that accesses files instead runs in a separate server process. While the client's approval prompt gates execution based on declared arguments, the spawned server process executes with ambient host filesystem permissions. An ostensibly benign operation can therefore access and expose host credentials through ordinary tool output. We call this vulnerability class \emph{filesystem capability laundering}: an out-of-scope filesystem capability is exercised under the guise of an innocuous tool contract.

We examine the gap at three layers. At the \emph{server-category} layer (L1), we compare inferred filesystem capability flags against the server's functional classification (\texttt{primary\_type}). Because non-filesystem servers (e.g., memory or code development) often perform legitimate local persistence, category mismatches serve as coarse screening signals. At the \emph{tool-name} layer (L2), a tool's narrow operation can reach files beyond its intended scope: \texttt{git\_add} can stage an out-of-repository file whose contents a later diff returns (CVE-2026-27735, \S\ref{sec:cases:gitadd}). At the \emph{defense-bypass} layer (L3), an existing path check can be defeated: the Filesystem MCP Server's string-prefix check admitted sibling directories sharing the allowed prefix (CVE-2025-53110~\cite{cve-fs-prefix}). Figure~\ref{fig:attack-dataflow} illustrates the staging-and-diff disclosure sequence.

We build MCP-FSD (\textbf{F}ilesystem \textbf{S}cope \textbf{D}etection), a framework that catalogs MCP servers across registries, discovers each server's tools inside per-repository sandboxes, screens candidates with MCP-aware static filters, and dynamically validates each one against an inferred least-privilege boundary. Across a catalog of 21{,}097 servers from nine registries, 787 of 4{,}806 triaged servers (16.4\%) receive a filesystem capability flag without a FileSystem category label. Dynamic validation yields 150 adjudicated instances under tested policies. We select 13 additional findings for coordinated disclosure and document two Git MCP Server CVEs (\S\ref{sec:cases}).

We make three contributions:
\begin{itemize}
  \item We characterize and formalize how attacker-influenced MCP tool arguments expose filesystem capabilities beyond a tool's contract and deployment scope.
  \item We design MCP-FSD to connect source screening and inferred per-tool boundaries to executable evidence, evaluating exploitability, benign operability, and sandbox containment across three complementary dynamic validation cases.
  \item We measure ecosystem-scale screening signals, confirm 150 vulnerable instances through dynamic validation under tested sandbox policies, and present Git case studies showing how repository operations expose host credentials.
\end{itemize}

We evaluate these contributions through five research questions (\S\ref{sec:evaluation}). \textbf{RQ1}: how frequent are screening signals, and what evidence supports them? \textbf{RQ2}: how many candidates survive three-case dynamic validation? \textbf{RQ3}: what were the outcomes of coordinated disclosure? \textbf{RQ4}: how do violations distribute over language, registry, and tool category? \textbf{RQ5}: how often does a confirmed instance need a format-wrapped payload rather than a bare traversal string?

\section{Background and Threat Model}
\label{sec:background}

\subsection{Threat Model}
\label{sec:background:threat}

\begin{figure}[t]
\centering
\includegraphics[width=\textwidth]{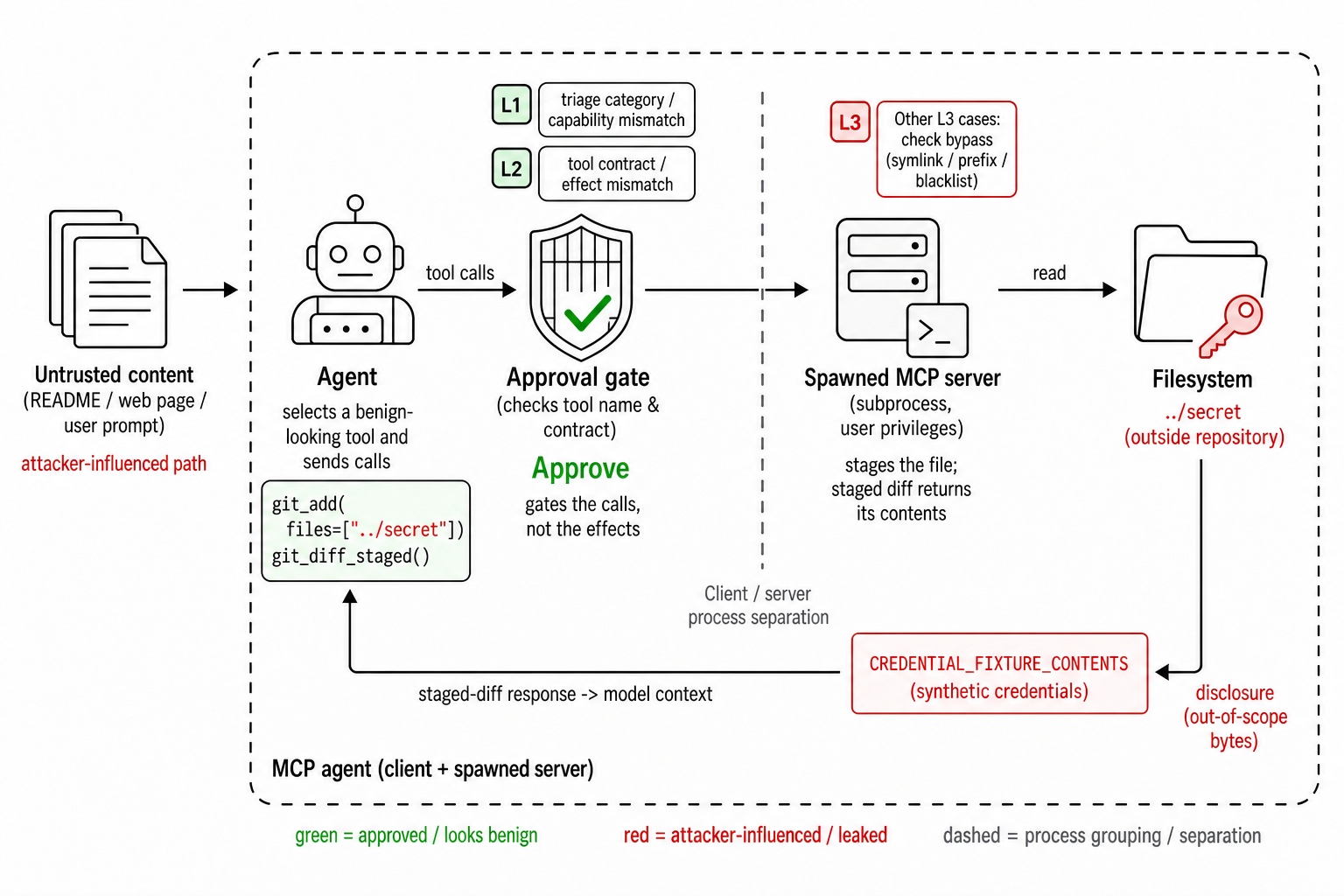}
\caption{Disclosure through a staging-and-diff sequence in the studied local-server deployment (\S\ref{sec:cases:gitadd}). The repository path is fixed for both calls, while the attacker-influenced file path selects a fixture outside it. Tool approval admits the calls, while the server's filesystem permissions allow the excess read. The response follows the second call, not staging alone.}
\label{fig:attack-dataflow}
\end{figure}

We study local stdio MCP servers~\cite{mcp-transports} launched as separate processes. In typical local deployments, the server runs with the user's ambient filesystem permissions, and the client's tool-approval prompt enforces no per-tool path restrictions on the underlying process~\cite{claude-code-mcp,gemini-cli-mcp,codex-sandbox}. Process isolation alone does not restrict access to the host filesystem unless explicit sandbox policies are applied. Figure~\ref{fig:attack-dataflow} traces the resulting dataflow, from the content that carries the adversary's argument to the file bytes that leave in the response.

The adversary influences the value of an argument that drives a side effect, through content the agent ingests as untrusted input~\cite{indirect-pi,injecagent,pi-tool-selection}. The agent supplies it to a tool whose contract appears benign, and the spawned server exercises a host capability outside what that contract implies. We measure filesystem disclosure: an attack succeeds when the MCP response contains the contents of a file outside the tool's legitimate scope. The evaluation begins at tool invocation and assumes that the agent issues the tested call or sequence and that the server has host permissions to access the target file. Arbitrary server-code modification and compromise of the OS sandbox are outside this model.

\subsection{Filesystem Capability Laundering}
\label{sec:background:definition}

An MCP tool exhibits \emph{filesystem capability laundering} when its declared contract does not cover the filesystem effect its arguments can steer. Three conditions characterize the mechanism: (i)~an implemented effect exceeds the scope justified by the tool contract and deployment, (ii)~attacker-influenced arguments select the filesystem target, and (iii)~the call is admitted while the surrounding controls fail to constrain that effect.

The three layers describe different aspects of this mechanism: L1 supplies candidates through category/capability screening, L2 identifies the per-tool contract/effect gap, and L3 identifies a defeated implementation control. They are not mutually exclusive: a missing or bypassable path check can implement a per-tool scope violation.

Prompt injection can influence the argument passed to a tool by affecting the model's instruction-following~\cite{pi-apps,formalizing-pi}. In contrast, capability laundering centers on the resulting server-side effect. Missing validation or flawed path checks (e.g., path traversal or symlink resolution) provide the underlying implementation mechanisms~\cite{pathtraversal-github,untrustide}.

\subsection{Problem Formalization}
\label{sec:background:formal}

All effects and responses below refer to a fixed deployment context $d$, specified by the server version, launch configuration, intended workspace, and initial filesystem state. Let $s$ be an MCP server and $T(s)$ its tools. Invoking a tool $t \in T(s)$ with an argument vector $a$ produces a set of filesystem effects and returns a response $R(t,a)$ to the client:
\[
  E(t,a) \;\subseteq\; \mathcal{F} \times \{\mathsf{read}, \mathsf{write}\},
\]
Here $\mathcal{F}$ denotes filesystem targets resolved at access time. Let $\mathcal{A}(t)$ denote the argument vectors an adversary can steer through content the agent ingests (\S\ref{sec:background:threat}), and let $\mathrm{Reach}(t) = \bigcup_{a \in \mathcal{A}(t)} E(t,a)$ collect everything those arguments can touch.

Let $C(t) = (\textit{name}, \textit{description}, \textit{schema})$ be the declared contract, and let $B(t;d) \subseteq \mathcal{F} \times \{\mathsf{read},\mathsf{write}\}$ denote the permissions justified by this contract and deployment context, including documented roots and legitimate runtime requirements. This \emph{implied scope} differs between a generic file tool authorized for arbitrary paths and a repository tool restricted to a selected workspace.

\smallskip\noindent\textbf{Definition 1 (Filesystem capability laundering).} A tool $t$ launders filesystem capability when $\mathrm{Reach}(t) \not\subseteq B(t;d)$.

\smallskip\noindent\textbf{Definition 2 (Disclosure witness).} A pair $(a^\star, f^\star)$ witnesses laundering in $t$ when $(f^\star, \mathsf{read}) \in E(t, a^\star) \setminus B(t;d)$ and the contents of $f^\star$ appear verbatim in $R(t, a^\star)$. For an explicitly stated call sequence $\pi$, we use combined effects $E(\pi)$ and response transcript $R(\pi)$, as in staging followed by a diff.

\smallskip
MCP-FSD computes a policy hypothesis $\widehat{B}(t;d)$ from read-write and read-only path sets $(\widehat{A}_{rw},\widehat{A}_{ro})$ (Algorithm~\ref{alg:boundary}). Reads are allowed in their union and writes in $\widehat{A}_{rw}$. A fixture outside this policy tests the witness predicate with $\widehat{B}$ substituted for $B$. We discuss the empirical characteristics and limitations of this inferred boundary in \S\ref{sec:discussion:limitations}.

\paragraph{Evidence for the intended boundary.}
Grounding an intended boundary $B(t;d)$ relies on a hierarchy of evidence. Explicit launch configurations, documented workspace roots, and developer-intended containment checks provide strong boundary signals. Intermediate signals, such as base-directory constants, indicate intended storage roots without enforcing permissions. Conversely, heuristic cues such as tool names or subprocess working directories provide weak intent and do not confine access. For each finding, the contract and deployment context must establish why the accessed target lies outside legitimate tool operation (\S\ref{sec:cases:synthesis}).

\subsection{Problem Statement and Challenges}
\label{sec:background:challenges}

Five challenges guide the design of MCP-FSD (\S\ref{sec:design}).

\textit{C1: Large, heterogeneous ecosystem.} MCP servers are distributed across registries, languages, and packaging conventions. Tool discovery requires source collection, language-specific installation, and transport handling before candidates can enter a common validation pipeline.

\textit{C2: Implicit filesystem boundaries.} A path parameter indicates where an argument can point, but does not establish which locations are authorized. MCP-FSD combines the declared contract with deployment evidence, such as a selected repository root, to construct the policy hypothesis used during testing.

\textit{C3: Library-mediated accesses.} Tool handlers often delegate filesystem operations to third-party libraries. For example, \texttt{git\_add} passes its file list to GitPython. Source review identifies this call, while dynamic testing determines whether the resulting contents reach an MCP response.

\textit{C4: Semantically diverse inputs.} A tool may accept a single path, a list of files, or a structured document containing a reference. Payload generation must synthesize schema-valid inputs so execution reaches the underlying filesystem operation.

\textit{C5: Connecting filesystem analysis to MCP.} Filesystem analysis~\cite{android-fs} must identify the callable MCP tools behind each finding and how their effects become observable. MCP-FSD uses tool associations, schemas, response construction, and stdio execution to connect source-level candidates to reproducible calls.

\section{MCP-FSD Design}
\label{sec:design}
\begin{figure}[t]
\centering
\includegraphics[width=\textwidth]{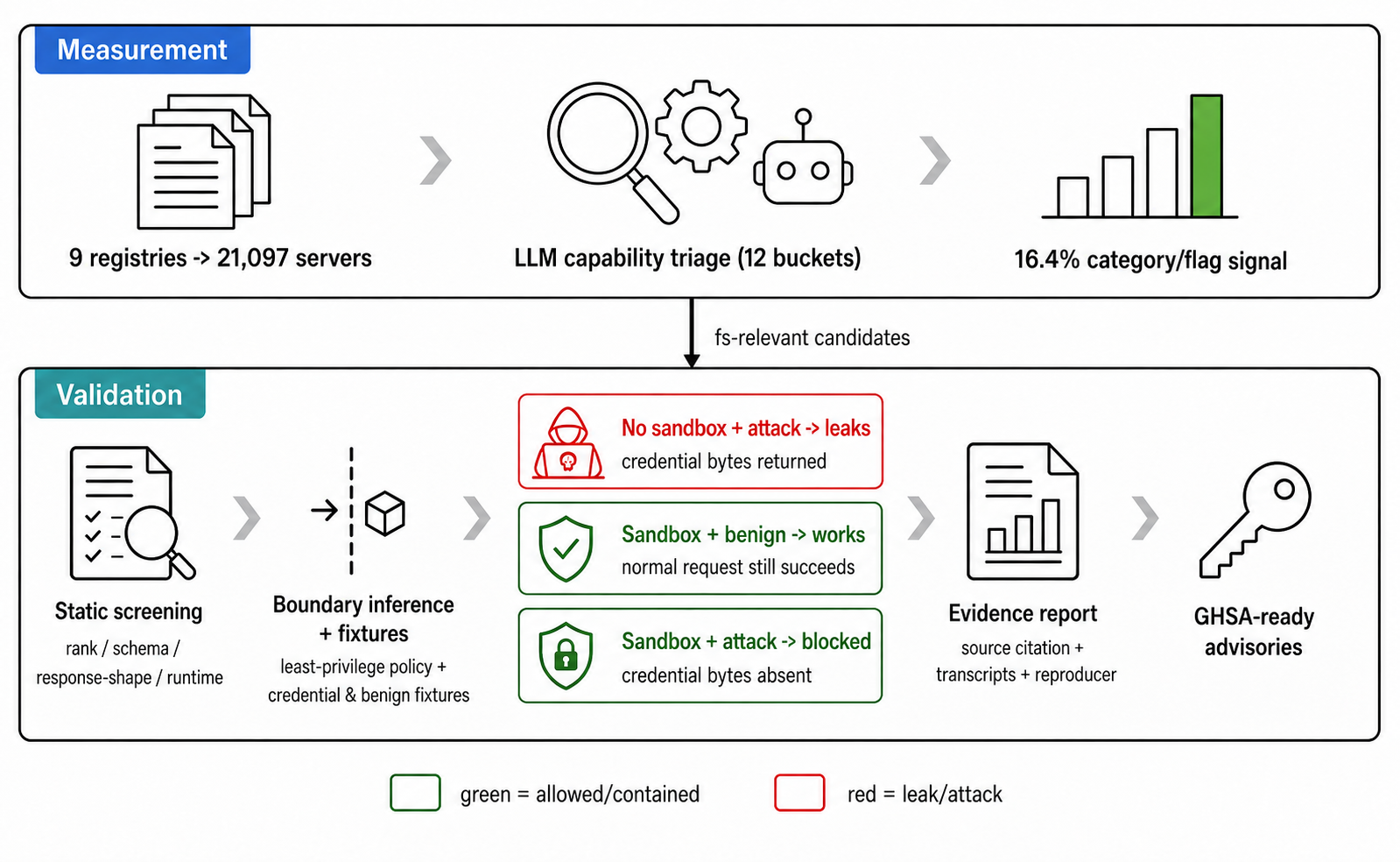}
\caption{MCP-FSD as two coupled passes: measurement builds and triages the MCP
corpus, while validation applies static screening, boundary inference,
generated payloads, and three response-body checks to produce evidence
reports.}
\label{fig:overview-mv}
\end{figure}

MCP-FSD runs in two passes (Figure~\ref{fig:overview-mv}). The \emph{measurement} pass crawls registries, discovers each server's tools and schemas, and assigns capability-triage labels. The \emph{validation} pass applies four MCP-aware static filters (\S\ref{sec:design:static}), infers a least-privilege boundary per tool (\S\ref{sec:design:boundary}), and confirms exploitability with three-case dynamic validation against that boundary (\S\ref{sec:design:dynamic}). The primary disclosure oracle checks for fixture contents in the MCP response. Selected instances use other transcript evidence (\S\ref{sec:eval:rq2}).

\subsection{Static Screening}
\label{sec:design:static}

MCP-FSD uses an LLM to identify tool registrations, handlers, and path parameters. It examines how these parameters reach filesystem operations and whether the code validates them. MCP-FSD records each candidate in the Static Analysis Results Interchange Format (SARIF), with its rule, source location, affected tools, parameter, sink, and rationale. We classify each finding into five rule categories spanning CWE-22 (path traversal) and CWE-59 (link resolution): \emph{PT-001} missing validation, \emph{PT-002} prefix-match containment, \emph{PT-003} URL traversal, \emph{PT-004} symlink TOCTOU (time-of-check-to-time-of-use), and \emph{PT-005} blacklist filter. A small long tail of specialty rules that a language model proposes is reported separately.

The scan reviews the server repository and can flag visible library calls for dynamic testing without reconstructing their effects. It does not exhaustively analyze dependency internals. We call one SARIF result at a reported fix site an \emph{emission}. A finding in a shared helper may affect several tools but produce only one emission. Emissions and dynamically confirmed instances therefore use different counting units.

Four MCP-aware filters select candidates for dynamic testing based on whether they can plausibly return out-of-scope file contents. \emph{M1} ranks SARIF findings tagged to MCP tools~\cite{llmxcpg}. \emph{M2} reads the tool schema for path-shaped parameters. \emph{M3} inspects response construction for a plausible file-content return. \emph{M4} keeps only candidates that run as stdio MCP servers. These lightweight filters prioritize high-probability candidates for dynamic testing without the overhead of whole-program taint tracking.

For the disclosure-oriented pass, M1 prioritizes file-content sinks over write-only or shell findings. M2 identifies the argument that reaches the sink, including entries in a tool's file list. M3 distinguishes a response containing file text from one reporting only derived metadata: a line count or schema descriptor cannot satisfy the credential-content oracle. M4 checks whether the candidate can be exercised in the local stdio harness. It excludes hosted services whose runtime cannot be reproduced.

\paragraph{Worked screening example.}
For \texttt{git\_add}, the reviewed handler connects the schema's \texttt{files} parameter to \texttt{repo.index.add(files)} without a repository-containment check, recorded as a PT-001 SARIF candidate. M1 associates it with the MCP tool and M2 identifies its path input. The disclosure observation comes from \texttt{git\_diff\_staged}, so M3 uses the explicitly stated staging-and-diff sequence rather than treating a staging acknowledgment as file content. M4 checks the stdio launch. While this evaluation validates this specific multi-call chain, dynamic validation confirms whether the final response discloses fixture contents.

\subsection{Boundary Inference Algorithm}
\label{sec:design:boundary}

For each candidate that passes static screening, boundary inference constructs the least-privilege scope used by payload generation and sandbox validation. MCP-FSD derives the scope from the tool's legitimate operation rather than from the suspected bug: read-write access covers the working directory, while read-only access covers the server binary, dependency tree, and required system-root subtree. The fixture planner then places the credential fixture outside every allow scope and the benign marker inside the working scope.

\paragraph{Algorithm.}
Algorithm~\ref{alg:boundary} uses four signals: launch configuration $L$, path-bearing schema fields $S$, the tool description $D$, and repository source signals $R$. The language model drafts a candidate policy from $(L,S,D)$. MCP-FSD extends the policy to cover runtime requirements identified in $R$, then tests whether a benign invocation succeeds.

\begin{algorithm}[t]
\caption{Boundary inference for an MCP tool $t$.}
\label{alg:boundary}
\begin{algorithmic}[1]
\Require launch tuple $L$, schema $S$, description $D$, repo signals $R$
\Ensure scope $(\widehat{A}_{rw}, \widehat{A}_{ro})$ as glob sets
\State $L_{\text{root}} \gets$ \Call{ExtractRoots}{$L$} \Comment{cwd, \texttt{--root}, env path roots}
\State $S_{\text{path}} \gets$ \Call{PathParams}{$S$} \Comment{parameters whose type or name encodes a path}
\State $R_{\text{static}} \gets$ \Call{HandlerPathRefs}{$R$} \Comment{source-visible paths and modes}
\State $\mathcal{P} \gets$ \Call{LMDraftPolicy}{$L_{\text{root}}, S_{\text{path}}, D$} \Comment{draft candidate scope}
\If{$R_{\text{static}} \not\subseteq \mathcal{P}$}
  \State $\mathcal{P} \gets \mathcal{P} \cup$ \Call{Cover}{$R_{\text{static}}$} \Comment{cover identified runtime needs}
\EndIf
\State \Return \Call{SplitRWRO}{$\mathcal{P}, S$} \Comment{partition by per-parameter read/write evidence}
\end{algorithmic}
\end{algorithm}

\paragraph{Subroutines.}
\textsc{ExtractRoots} reads the launch tuple: the working directory the client starts the server in, any \texttt{-{}-root}-style launch argument, and path-valued environment variables. \textsc{PathParams} walks the JSON Schema and returns the fields whose type, name, or format encodes a path, so payload generation targets the appropriate argument (C4). \textsc{HandlerPathRefs} identifies literal and constructed paths and access modes in the reviewed handler code. \textsc{LMDraftPolicy} asks the model for a candidate glob policy from those roots, fields, and the description. \textsc{Cover} extends the policy to include identified runtime dependencies (such as language runtimes, libraries, and system configuration paths). \textsc{SplitRWRO} partitions the result into a read-write and a read-only half according to whether each parameter is written or only read at the observed sites.

\paragraph{Checks and example.}
Before dynamic testing, MCP-FSD checks that the policy covers identified runtime requirements and leaves a non-empty out-of-scope region for the fixture. For \texttt{git\_add} this yields \texttt{/ws/**} read-write and dependencies read-only, with the benign file inside \texttt{/ws} and the credential fixture outside. The three-case test then replays staging followed by a diff (\S\ref{sec:cases:gitadd}).

The repository needs write access for the index and object-database updates required by staging, while runtime dependencies need only be readable. The credential fixture must lie outside both sets: placing it outside the working tree alone would be insufficient if a read-only runtime allowance still covered it. The benign invocation checks that these permissions support the intended staging-and-diff operation.

\subsection{Payload Generation and Dynamic Three-Case Validation}
\label{sec:design:dynamic}

\paragraph{Dynamic payload generation.}
For each path-shaped parameter, MCP-FSD first fixes two targets from the inferred boundary: the malicious target is the planted out-of-scope credential fixture, and the benign target is the in-scope marker file. A language model then proposes concrete argument forms, seeded with published path-traversal exemplars, and refines them against the tool schema, the inferred boundary, and observed responses until the malicious attempt reaches the fixed credential target or the attempt budget is exhausted. For a tool that parses structured input, it wraps the same target in a format-valid container, for example a BPMN, glTF, DOCX, or PNG file, or a Terraform body. The model varies the argument form while the target stays fixed. MCP-FSD resets test state and replays the same converged payload or stated call sequence with and without the sandbox.

Response-disclosure validation uses three cases. Other effects require separate evidence rules (\S\ref{sec:eval:rq2}). PoC~1 runs the malicious payload without the sandbox and succeeds only when the response contains the planted credential contents (C2--C3). PoC~2 runs the benign in-scope payload with the sandbox and requires the expected successful operation with no denial event. PoC~3 replays PoC~1's malicious payload with the sandbox and requires absent credential contents plus an explicit denial or a non-disclosing response.

This three-case protocol isolates exploitability from containment: PoC~1 confirms that the exploit succeeds on an unconfined host; PoC~2 verifies that the inferred sandbox policy preserves legitimate, in-scope functionality; and PoC~3 demonstrates that the sandbox neutralizes the exploit under that exact policy. Together, these cases establish that detected disclosures represent genuine vulnerabilities remediated by least-privilege containment.

We distinguish explicit containment, where the response exposes the operating-system denial, from silent containment, where the response is empty or non-disclosing without an exposed denial. A silent response alone does not identify which syscall was blocked. Throughout, we call this enforcement \emph{operating-system process isolation}, reserving ``kernel-enforced'' for backends that provide kernel-level enforcement such as Landlock or Seatbelt~\cite{nono,sandbox-runtime,ebpf-syscall}.
All 150 dynamically adjudicated instances in \S\ref{sec:eval:rq2} pass this three-case gate, providing the containment evidence discussed in \S\ref{sec:discussion:mitigations}.

Per candidate, MCP-FSD records the three parsed response transcripts and the reproducer script, along with source citations, the inferred sandbox policy, runtime details and, when available, a sink citation pinned to a specific file and line. The verdict adjudicator reads this per-candidate record.

\section{Evaluation}
\label{sec:evaluation}

\subsection{Experiment Setup}
\label{sec:eval:setup}

\paragraph{Data collection.} We catalog 21{,}097 listings from nine registries, including 17{,}919 open-source entries. The measurement pass selects local open-source servers, excluding forks, archived repositories, remote-hosted services, and unavailable code. Successful startup and discovery produce the triage input. The validation pass then screens and tests selected candidates.

The triage step classifies each server across twelve functional categories (\texttt{primary\_type}), such as FileSystem, CodeDev, and KnowledgeRetrieval, derived via LLM analysis of tool descriptions rather than external registry tags.

\paragraph{Implementation and corpus funnel.} MCP-FSD comprises 7{,}486 executable lines of Python and shell plus 618 lines of analysis prompts. Dynamic validation runs each candidate in \texttt{nono}~v0.43.0~\cite{nono}, a macOS Seatbelt sandbox. Static analysis returns results for 4{,}292 servers and flags 2{,}818 of them for potential vulnerabilities (20.3\% and 13.4\% of the 21{,}097-server catalog, respectively).
\paragraph{Units of analysis.}
Our evaluation distinguishes four units of analysis: \emph{servers} (catalog and triage entries); \emph{static emissions} (individual SARIF finding sites, where a shared helper may affect multiple tools); \emph{dynamic instances} (unique tool-policy pairs confirmed via three-case replay); and \emph{disclosure packages} (curated multi-tool findings bundled for vendor remediation).

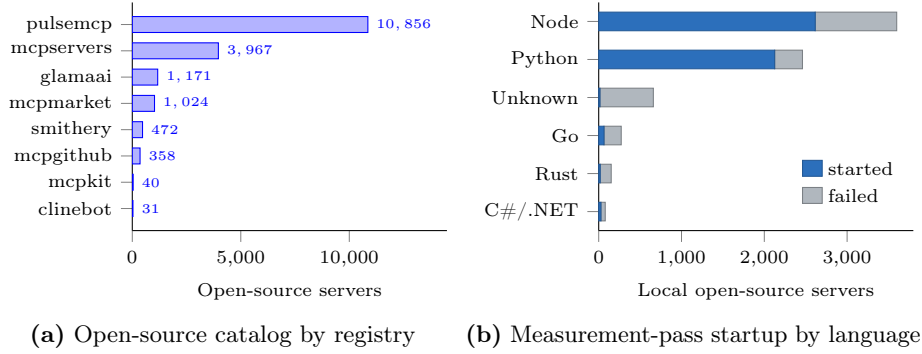
\begin{figure}[t]
\centering
\begin{minipage}[t]{0.49\linewidth}
\centering
\begin{tikzpicture}
\begin{axis}[distbar, width=0.96\linewidth, height=4.6cm, xmax=14500,
  bar width=6pt,
  scaled x ticks=false,
  xlabel={Open-source servers}, xlabel style={font=\scriptsize},
  yticklabel style={font=\scriptsize}, xticklabel style={font=\scriptsize},
  every node near coord/.append style={font=\tiny},
  symbolic y coords={pulsemcp, mcpservers, glamaai, mcpmarket, smithery, mcpgithub, mcpkit, clinebot},
  nodes near coords={\pgfmathprintnumber[int detect,1000 sep={,}]{\pgfplotspointmeta}},
]
\addplot coordinates {(10856,pulsemcp) (3967,mcpservers) (1171,glamaai) (1024,mcpmarket) (472,smithery) (358,mcpgithub) (40,mcpkit) (31,clinebot)};
\end{axis}
\end{tikzpicture}
\\[2pt]
{\footnotesize\textbf{(a)} Open-source catalog by registry}
\end{minipage}\hfill
\begin{minipage}[t]{0.49\linewidth}
\centering
\begin{tikzpicture}
\begin{axis}[
  xbar stacked, y dir=reverse,
  width=0.96\linewidth, height=4.6cm, bar width=7pt,
  xmin=0, xmax=3800,
  axis x line*=bottom, axis y line*=left, tick align=outside,
  ytick=data,
  symbolic y coords={Node, Python, Unknown, Go, Rust, {C\#/.NET}},
  yticklabel style={font=\scriptsize}, xticklabel style={font=\scriptsize},
  xlabel={Local open-source servers}, xlabel style={font=\scriptsize},
  legend style={font=\scriptsize, at={(0.98,0.06)}, anchor=south east, draw=none, fill=none},
  legend cell align=left,
]
\addplot[fill=distbarmain, draw=distbarmain!80!black] coordinates {(2615,Node) (2125,Python) (15,Unknown) (64,Go) (18,Rust) (27,{C\#/.NET})};
\addplot[fill=distbaralt, draw=distbaralt!70!black] coordinates {(984,Node) (336,Python) (645,Unknown) (208,Go) (133,Rust) (52,{C\#/.NET})};
\legend{started, failed}
\end{axis}
\end{tikzpicture}
\\[2pt]
{\footnotesize\textbf{(b)} Measurement-pass startup by language}
\end{minipage}
\caption{Corpus composition: (a)~17{,}919 open-source listings by registry. (b)~Startup outcomes for 7{,}222 local open-source servers, by language: 86.3\% of Python and 72.7\% of Node servers start, against 34.2\% (C\#/.NET), 23.5\% (Go), and 11.9\% (Rust).}
\label{fig:corpus}
\end{figure}
\paragraph{Triage snapshot and selection bias.} The final triage snapshot contains 4{,}806 servers, with static analysis evaluated on the successfully cloned and parsed subset. The funnel is selective: only 7{,}222 of 21{,}097 catalogued servers (34.2\%) reach the measurement pass, the post-measurement corpus is 97\% Node and Python, and pulsemcp supplies 84\% of measurement-pass startups. Our prevalence estimates and per-registry comparisons describe this locally runnable open-source corpus rather than an idealized, uniform distribution across all ecosystems.

\subsection{RQ1: Screening Signals and Evidence}
\label{sec:eval:rq1}

The L1 signal combines the triage's multi-label \texttt{primary\_type} field and four filesystem capability flags (file-related call, path input, file read, and file write), extracted via LLM analysis of tool discovery summaries. Among 4{,}806 servers, 787 (16.4\%) exhibit at least one filesystem capability flag despite lacking a FileSystem category classification. Because routine persistence mechanisms do not necessarily violate a tool's contract, and absent external ground-truth labels across the entire catalog, this category mismatch serves as a coarse initial screening signal rather than a definitive violation metric.

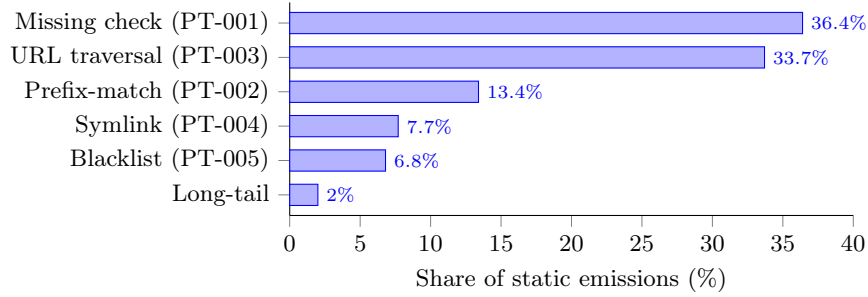
\begin{figure}[tbp]
\centering
\begin{tikzpicture}
\begin{axis}[distbar, width=0.74\linewidth, height=4.4cm, xmax=40,
  xlabel={Share of static emissions (\%)},
  symbolic y coords={{Missing check (PT-001)},{URL traversal (PT-003)},{Prefix-match (PT-002)},{Symlink (PT-004)},{Blacklist (PT-005)},{Long-tail}},
  nodes near coords={\pgfmathprintnumber[fixed,precision=1]{\pgfplotspointmeta}\%},
]
\addplot coordinates {(36.4,{Missing check (PT-001)}) (33.7,{URL traversal (PT-003)}) (13.4,{Prefix-match (PT-002)}) (7.7,{Symlink (PT-004)}) (6.8,{Blacklist (PT-005)}) (2.0,{Long-tail})};
\end{axis}
\end{tikzpicture}
\caption{Failure-mode distribution over 10{,}304 static emissions from 2{,}818 flagged servers.}
\label{fig:failuremodes}
\end{figure}

At the static-emission level, MCP-FSD emits 10{,}304 findings across the 2{,}818 flagged servers (65.7\% of the 4{,}292 with results), with missing path validation as the largest individual rule category (Figure~\ref{fig:failuremodes}). Combined, the defense-bypass rules (PT-002--PT-005, 61.6\%) outweigh PT-001 missing-validation (36.4\%). Rule families rank differently when counted by confirmed instances rather than static emissions: a single repository can emit many findings across sibling tools, increasing the weight of whichever rule it carries. Each emission has a severity level: 61.8\% are reported at \texttt{error} level, 29.8\% at \texttt{warning}, and 8.4\% at \texttt{note}.

URL-traversal emissions include candidates that do not disclose local files, so static counts remain separate from dynamic outcomes.

\paragraph{Layer-classifier accuracy.} Because the L2 and L3 results rest on automated classification, we validate each classifier against a hand-labeled sample of about 300 candidates (Table~\ref{tab:classifier-accuracy}). The L2 tool-name classifier reaches 99.5\% precision and 97.1\% recall over 300 labeled tools (one false positive, six false negatives). The L3 defense-bypass classifier achieves 95.3\% precision and 99.3\% recall over 297 labeled findings (seven false positives, one false negative). The remaining three repositories in the 300-sample set were excluded because they were deleted, made private, or inaccessible at validation time. These benchmarks measure static classification precision; end-to-end authorization fidelity is evaluated dynamically in RQ2.

\begin{table}[tbp]
\centering
\caption{Per-layer detection accuracy. Pos./Neg.\ are ground-truth positive/negative counts on the hand-labeled, stratified benchmark for each layer (300 for L2, 297 for L3). L1 accuracy is unmeasured because no independent labeled benchmark exists.}
\label{tab:classifier-accuracy}
\small
\begin{tabular*}{\textwidth}{@{\extracolsep{\fill}}lrrrrrrrr@{}}
\toprule
Layer & Pos. & Neg. & TP & FP & FN & Prec. & Rec. & F1 \\
\midrule
L1 server-category & -- & -- & -- & -- & -- & -- & -- & -- \\
L2 tool-name       & 205 & 95  & 199 & 1 & 6 & 0.995 & 0.971 & 0.983 \\
L3 defense-bypass  & 142 & 155 & 141 & 7 & 1 & 0.953 & 0.993 & 0.972 \\
\midrule
\textbf{Micro} & 347 & 250 & 340 & 8 & 7 & 0.977 & 0.980 & 0.978 \\
\textbf{Macro} & -- & -- & -- & -- & -- & 0.974 & 0.982 & 0.978 \\
\bottomrule
\end{tabular*}
\end{table}

\subsection{RQ2: Confirmed Instances}
\label{sec:eval:rq2}

MCP-FSD confirms 150 instances, each adjudicated under the three-case dynamic validation (\S\ref{sec:design:dynamic}) against the inferred least-privilege boundary. For 136 of the 150 instances, adjudication relies on exact credential-marker matching: PoC~1's response contains the planted secret string, whereas PoC~3's response contains no marker bytes. For the remaining 14, the adjudicator evaluates corresponding evidence from other transcript features (write side-effects, response-shape changes, or external-CLI outputs).

\subsection{RQ3: Disclosed Outcomes}
\label{sec:eval:rq3}

We reported 13 findings, including two CVEs in Anthropic's Git MCP Server (\S\ref{sec:cases}). MCP-FSD dynamically validates \texttt{git\_add} with a full three-case PoC. The \texttt{git\_init} finding is static because its disclosure is indirect: it seeds a repository that downstream Git tools then read. Both CVEs have been patched upstream.

Among the 13 packages, 11 cite source-level mechanisms or configuration as evidence, including validators, storage-base constants, and shell argument handling. These evidence sources span a spectrum of enforcement: a base directory may lack runtime checks, while an overbroad allowlist or shell-quoting flaw exhibits a distinct authorization boundary from a containment bypass. Two packages rely on contract-implied scope boundaries without explicit path validation at the read sink. All 13 packages were submitted for coordinated disclosure, with upstream fixes or CVE assignments pending at the reporting cutoff.

We prioritized findings where either a source-level control is bypassed or the project publicly commits to a narrow scope that the implementation does not enforce, because those reports have a per-server fix without requiring a protocol change.

\subsection{RQ4: Violation Distribution}
\label{sec:eval:rq4}
\begin{figure}[tbp]
\centering
\begin{minipage}[t]{0.49\linewidth}
\centering
\begin{tikzpicture}
\begin{axis}[
  xbar stacked, y dir=reverse,
  width=\linewidth, height=4.4cm,
  bar width=6pt,
  xmin=0, xmax=100,
  axis x line*=bottom, axis y line*=left, tick align=outside,
  ytick=data,
  symbolic y coords={
    {PT-001 (no check)},
    {PT-002 (prefix-match)},
    {PT-003 (URL traversal)},
    {PT-004 (symlink)},
    {PT-005 (blacklist)},
  },
  yticklabels={PT-001,PT-002,PT-003,PT-004,PT-005},
  yticklabel style={font=\scriptsize},
  xticklabel style={font=\scriptsize},
  xlabel={Share of rule's emissions (\%)},
  xlabel style={font=\scriptsize},
  legend style={font=\scriptsize, at={(0.5,0)}, anchor=north, yshift=-35pt,
                draw=none, fill=none, legend columns=-1,
                /tikz/every even column/.append style={column sep=4pt}},
  legend cell align=left,
  enlarge y limits=0.10,
]
\addplot+[fill=blue!55,  draw=blue!70!black]
  coordinates {(43.9,{PT-001 (no check)}) (59.0,{PT-002 (prefix-match)})
               (58.6,{PT-003 (URL traversal)}) (57.5,{PT-004 (symlink)})
               (63.0,{PT-005 (blacklist)})};
\addplot+[fill=orange!65, draw=orange!75!black]
  coordinates {(53.8,{PT-001 (no check)}) (38.2,{PT-002 (prefix-match)})
               (39.7,{PT-003 (URL traversal)}) (36.8,{PT-004 (symlink)})
               (32.1,{PT-005 (blacklist)})};
\addplot+[fill=green!55,  draw=green!70!black]
  coordinates {(0.7,{PT-001 (no check)}) (1.6,{PT-002 (prefix-match)})
               (0.5,{PT-003 (URL traversal)}) (2.8,{PT-004 (symlink)})
               (3.2,{PT-005 (blacklist)})};
\addplot+[fill=gray!50,   draw=gray!70!black]
  coordinates {(1.6,{PT-001 (no check)}) (1.2,{PT-002 (prefix-match)})
               (1.2,{PT-003 (URL traversal)}) (2.9,{PT-004 (symlink)})
               (1.7,{PT-005 (blacklist)})};
\legend{Node, Python, Go, other}
\end{axis}
\end{tikzpicture}
\\[2pt]
{\footnotesize\textbf{(a)} Rule $\times$ language (within-rule share)}
\end{minipage}\hfill
\begin{minipage}[t]{0.49\linewidth}
\centering
\begin{tikzpicture}
\begin{axis}[
  width=\linewidth, height=4.4cm,
  xmin=0, xmax=100, ymin=0, ymax=100,
  xlabel={Bottom-X\% of flagged repos},
  ylabel={Cumulative \% of emissions},
  xlabel style={font=\scriptsize}, ylabel style={font=\scriptsize, yshift=-2pt},
  xticklabel style={font=\scriptsize}, yticklabel style={font=\scriptsize},
  grid=major, grid style={dotted,gray!40},
  axis x line*=bottom, axis y line*=left,
  legend style={font=\scriptsize, at={(0.5,1.03)}, anchor=south,
                draw=none, fill=none, legend columns=2},
  legend cell align=left,
]
\addplot[blue!70!black, line width=1pt, mark=*, mark size=1pt]
  coordinates {(0,0) (10,2.7) (25,8.0) (50,23.6) (75,47.5)
               (90,70.1) (95,81.2) (99,93.9) (100,100)};
\addlegendentry{Lorenz}
\addplot[dashed, gray!70, line width=0.6pt, no marks]
  coordinates {(0,0) (100,100)};
\addlegendentry{equality}
\node[anchor=west, font=\tiny, fill=white, inner sep=1pt]
  at (axis cs:6,67) {top 5\% $\Rightarrow$ 18.7\%};
\node[anchor=west, font=\tiny, fill=white, inner sep=1pt]
  at (axis cs:6,53) {top 25\% $\Rightarrow$ 52.5\%};
\end{axis}
\end{tikzpicture}
\\[2pt]
{\footnotesize\textbf{(b)} Emission concentration (Gini $\approx 0.55$)}
\end{minipage}

\caption{\textbf{Static findings by language and repository.} \textbf{(a)}~Python contributes the largest share of PT-001 emissions, while Node leads PT-002--PT-005. PT-003 subsumes URL-based access patterns. \textbf{(b)}~Across 2{,}818 flagged repositories, the top 5\% of repos produce 18.7\% of emissions and the top 25\% produce 52.5\%.}
\label{fig:rule-lang-lorenz}
\end{figure}
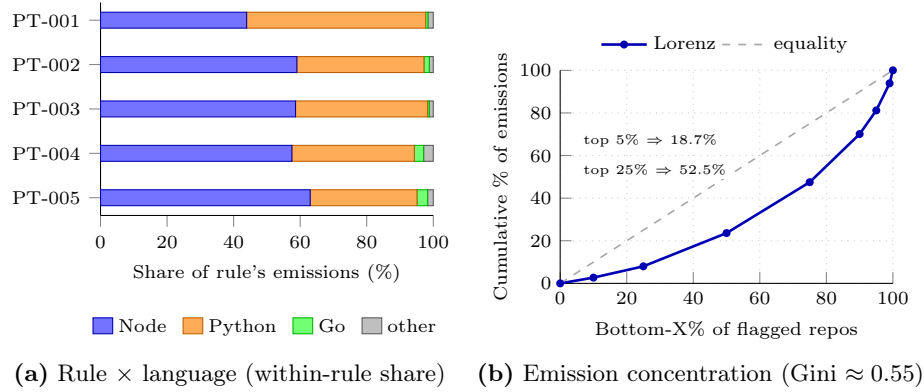

The flagged population follows the same concentration as the corpus: pulsemcp dominates the registry axis, and Node/Python dominate the language axis. At the sink level, 54.3\% of emissions are Node-ecosystem files and 42.6\% are Python files. All other languages together account for under 3.1\%. Python leads PT-001 missing-validation (53.8\% within-rule share), while Node leads every bypass-style rule at 57\%--63\% (Figure~\ref{fig:rule-lang-lorenz}a). This distribution reflects structural tooling gaps: the MCP TypeScript SDK provides no built-in path-containment primitive (such as a \texttt{safePath()} utility), leaving Node servers prone to flawed custom checks (accounting for 54.3\% of all emissions). Conversely, Python servers frequently omit validation checks entirely (53.8\% of PT-001). These patterns indicate distinct remediation paths: default allowlist helpers for Python and standardized path-containment primitives for TypeScript.

Table~\ref{tab:fs-footprint} summarizes the filesystem footprint across the 4{,}806 triaged servers. In total, 787 servers exhibit filesystem capability flags without a FileSystem category classification, forming the L1 screening candidate set.

\begin{table}[tbp]
\centering
\caption{Filesystem footprint in the final triage snapshot. The L1 row counts a positive filesystem flag without a FileSystem label.}
\label{tab:fs-footprint}
\small
\begin{tabular*}{0.82\linewidth}{@{\extracolsep{\fill}}lrr@{}}
\toprule
Measurement & Count & Share of triaged \\
\midrule
Triaged servers & 4{,}806 & 100.0\% \\
FileSystem label & 1{,}345 & 28.0\% \\
Any filesystem capability flag & 2{,}131 & 44.3\% \\
L1 screening signal & 787 & 16.4\% \\
\bottomrule
\end{tabular*}
\end{table}

\subsection{RQ5: Payload Forms}
\label{sec:eval:rq5}

We categorize the payloads synthesized by the iterative generation loop (\S\ref{sec:design:dynamic}) by their converged input format (bare traversal paths versus format-valid containers). Across the 150 confirmed instances, 144 (96.0\%) use the bare traversal form, while 6 (4.0\%) require a format-valid container. Among the 38 confirmed instances that bypass an explicit source-level validation check, 32 (84.2\%) use the bare form and the same 6 (15.8\%) require wrapping.

Format-valid fixtures ensure that tools parsing structured inputs (such as glTF, DOCX, or Terraform configurations) accept the payload before executing underlying filesystem calls. An unformatted traversal string rejected by an input parser would leave the tool's access to the target untested. While 96.0\% of vulnerable tools accept bare paths, structured tools necessitate format-valid containers to expose the laundering path.

\newpage
\noindent\begin{minipage}{\linewidth}
\section{Case Studies}
\label{sec:cases}

Two CVEs in Anthropic's Git MCP Server illustrate L2 laundering across distinct boundaries. MCP-FSD flags both and dynamically validates \texttt{git\_add} (\S\ref{sec:eval:rq3}).

\subsection{Case 1 (L2): Unscoped Repository Creation in \texttt{git\_init} (CVE-2025-68143)}
\label{sec:cases:gitinit}

The \texttt{git\_init} handler initializes a repository at the supplied \texttt{repo\_path}:

\begin{lstlisting}[style=paper]
# git_init(repo_path)
repo_path = Path(arguments["repo_path"])
repo = git.Repo.init(path=repo_path, mkdir=True)
\end{lstlisting}
\end{minipage}\par

\noindent\textbf{Root cause.} The handler passes the caller-supplied \texttt{repo\_path} directly to \texttt{Repo.init()}, without verifying containment within the configured workspace. Any writable filesystem location becomes a valid initialization target. \textbf{Impact.} When an attacker-influenced argument names a sensitive directory such as \texttt{\textasciitilde/.ssh}, \texttt{git\_init} converts that directory into a Git repository. Subsequent Git tool calls such as \texttt{git\_add} and \texttt{git\_diff\_staged} then operate inside it legitimately and return raw file contents, including SSH private keys, as diff output to the agent. In this deployment, the selected workspace defines the intended scope for repository creation. The issue was assigned CVE-2025-68143~\cite{cve-git-init}. The fix removes \texttt{git\_init} from the server (released as version \texttt{2025.9.25}). The upstream advisory credits Yarden Porat for the initial report and one of our co-authors (\texttt{@0dd}) for developing the upstream patch. We contributed independent vulnerability analysis and remediation.

Under this workspace boundary, the vulnerability satisfies all three capability-laundering conditions (\S\ref{sec:background:definition}): repository creation exceeds the authorized workspace scope (i), caller-supplied arguments steer the filesystem target (ii), and the server executes the initialization without containment (iii), enabling subsequent credential extraction.

\bigskip
\noindent\begin{minipage}{\linewidth}
\subsection{Case 2 (L2): Path-Traversal Staging in \texttt{git\_add} (CVE-2026-27735)}
\label{sec:cases:gitadd}

The \texttt{git\_add} handler stages the supplied \texttt{files} list:

\begin{lstlisting}[style=paper]
# git_add(repo_path, files=[...])
repo = git.Repo(repo_path)
repo.index.add(files)
\end{lstlisting}
\end{minipage}\par

\noindent\textbf{Root cause.} The handler does not check whether paths in \texttt{files} stay within the repository root before staging. The affected \texttt{\_to\_relative\_path()} function in GitPython returns relative paths without validating their containment. Paths containing \texttt{../} then reach \texttt{\_store\_path()}, which reads the file bytes into Git's object database. The contents are stored by Git even though the source file lies outside the working tree. The fix, released in \texttt{2026.1.14}, delegates staging to the Git CLI through \texttt{repo.git.add()}. The CLI rejects out-of-repository paths~\cite{cve-git-add}.

\noindent\textbf{Impact.} Sensitive host files such as SSH keys or dotfiles can be staged and then exposed through a staged diff, or carried into a commit that sends them to a remote on the next push.
The repository root forms the implied scope of \texttt{git\_add}; failing to enforce containment allows out-of-scope files to be staged into the repository index. The issue was assigned CVE-2026-27735~\cite{cve-git-add}. The complete three-case dynamic validation runs against the pre-patch version: the unsandboxed staging-and-diff sequence returns the planted credential contents, the sandboxed benign sequence succeeds on an in-scope file, and the sandboxed malicious sequence returns a permission-denied error (\texttt{EPERM}) with the credential bytes absent.

\subsection{Synthesis}
\label{sec:cases:synthesis}

The cases require containment at different points in the call sequence. For \texttt{git\_init}, the creation target must lie within the deployment's workspace before a repository is created there. Once a sensitive directory has become a repository, later calls that name the same \texttt{repo\_path} can stage and display its files without leaving that repository root. A check against only the newly selected root would therefore miss the earlier workspace violation. For \texttt{git\_add}, the repository already exists: each entry in \texttt{files} must remain within it. The preceding state-changing call makes the bytes reachable, and the staged-diff response reveals the disclosure.

In both cases, the intended scope, caller-selected path, and missing containment instantiate conditions (i)--(iii) of \S\ref{sec:background:definition}. Neither case requires a pre-existing containment check. The same reasoning applies to other tools: a configured vault root supports an intended scope, whereas a suggestive name such as \texttt{read\_build\_log} exhibits greater scope ambiguity.

Prior L3 vulnerabilities in the reference Filesystem MCP Server---such as sibling-prefix bypass (CVE-2025-53110~\cite{cve-fs-prefix}) and symlink traversal (CVE-2025-53109~\cite{cve-fs-symlink})---provide external benchmark cases. MCP-FSD flags their source-level patterns to verify static rule coverage, though they remain separate from our 13 newly reported findings.

\section{Discussion}
\label{sec:discussion}

\subsection{Mitigations}
\label{sec:discussion:mitigations}

At L3, developers should enforce canonical-path containment, validate schema fields, and address symlink races at access time. An SDK-level containment helper would reach the many Node servers that currently ship no such check (\S\ref{sec:eval:rq4}). At L2, a protocol extension could expose a per-tool \texttt{fs\_scope} field in \texttt{tools/list}, allowing clients to inspect the requested read/write scope before approval and enforce the approved policy on server execution~\cite{isolategpt}. MCP-FSD's read-write and read-only path sets offer a candidate representation for that field. Inferred policies would need review against deployment requirements. Our three-case validation demonstrates that operating-system process isolation reliably enforces least-privilege boundaries without disrupting benign tool functionality. At L1, registry screening can identify filesystem capabilities for further review.

\subsection{Limitations}
\label{sec:discussion:limitations}

First, the pipeline only validates servers it can install and run. The dynamic results are conditional on successful startup and the selected runtime configurations. Second, dynamic validation may miss a violation whose out-of-scope effect never appears in a response, or that requires a runtime precondition the harness does not reproduce.
Third, we do not measure the accuracy of the inferred boundaries themselves. A policy that is too restrictive can flag legitimate access, while one that is too permissive can miss a violation. Source coverage and a successful benign call do not establish authorization accuracy. Future benchmark efforts should establish expert-annotated boundaries across diverse server documentation to systematically quantify boundary inference precision and recall.

Fourth, the effect model includes writes, but the disclosure-oriented evaluation does not systematically measure unauthorized creation, modification, or deletion. The \texttt{git\_init} case illustrates write-driven setup, and our separate Memory MCP Server report demonstrates configuration injection through a persistence write when an attacker controls the launch-time persistence path. Its fix was merged upstream without a CVE.

Assessing these writes requires filesystem-state evidence as well as responses. In a targeted case study on \texttt{mcp-architect}, a caller-controlled \texttt{projectId} is joined to the storage root without containment: an unsandboxed invocation successfully writes a sentinel file outside the allowed storage directory, whereas sandboxed replay is blocked by the OS policy (\texttt{EPERM}). This demonstrates how an integrity-focused oracle can detect unauthorized filesystem writes even when no file contents are returned in tool responses.

Fifth, enforcement strength follows the sandbox backend (\S\ref{sec:design:dynamic}). Sixth, validation covers selected tools and explicitly stated call sequences, not systematic discovery of cross-tool chains. A tool that writes attacker-controlled content into a file that a second tool later reads can launder capability across the pair while each tool, evaluated alone, respects its boundary.

Seventh, several stages rely on LLM-generated judgments, namely the capability-triage labels used in the L1 signal (\S\ref{sec:eval:rq1}), the rule families behind the static emissions, and the boundary-policy drafts. Those results inherit the model's error modes. A systematic accuracy study of these model-dependent stages remains future work, along with full reporting of model versions, prompts, and decoding settings.

\subsection{Ethical Considerations}
\label{sec:discussion:ethics}

This work analyzes publicly available MCP servers and validates disclosure against fixture credentials, never real secrets. We follow responsible-disclosure practice: the Git MCP cases of \S\ref{sec:cases} were reported to the affected vendor, and both received CVE assignments. Dynamic validation runs on controlled installations with synthetic fixtures and resets between cases. PoC~1 intentionally omits the test sandbox.

\section{Related Work}
\label{sec:related}

\paragraph{MCP security and filesystem detection.} Table~\ref{tab:related} situates MCP-FSD among MCP security studies~\cite{micryscope,mcpguard,mcp-context-injection,mcp-sandboxscan,dci-shi-2026,mcp-landscape} and filesystem-flaw detectors~\cite{pathtraversal-github,untrustide,android-fs,llmxcpg}. The closest contract-analysis work, DCIChecker~\cite{dci-shi-2026}, combines structured code extraction with direct, reverse, and arbitration prompts to detect description-code inconsistencies. MCP-FSD focuses on filesystem effects triggered by attacker-controlled arguments and tests them against an inferred policy for each tool.

\begin{table}[tbp]

\centering
\caption{Related work by analysis target and validation evidence.}
\label{tab:related}
\small
\begin{tabularx}{\textwidth}{@{}>{\raggedright\arraybackslash}p{0.24\textwidth}>{\raggedright\arraybackslash}X>{\raggedright\arraybackslash}X@{}}
\toprule
Work & Target / scope & Evidence / method \\
\midrule
PathSentinel~\cite{android-fs} & Android filesystem attacks and platform access-control policy & Program and policy analysis with generated exploit validation \\
MICRYSCOPE~\cite{micryscope} & Cryptographic misuse in MCP servers & Cross-language API and dependency analysis with taint rules \\
MCPGuard~\cite{mcpguard} & MCP threats and defenses & Systematization of scanning and runtime-monitoring approaches \\
MCP-SandboxScan~\cite{mcp-sandboxscan} & External-input exposure and filesystem capability violations & WASM/WASI execution with output matching and denial evidence \\
DCIChecker~\cite{dci-shi-2026} & Description-code inconsistency, including undeclared effects & Structured code extraction and LLM cross-validation \\
MCP-FSD & Filesystem effects relative to per-tool policy hypotheses & Source screening and fixture-based three-case execution under OS containment \\
\bottomrule
\end{tabularx}

\end{table}

While MCP-SandboxScan~\cite{mcp-sandboxscan} utilizes WASI sandboxing and output matching to detect external-input leakage, MCP-FSD couples static screening with dynamic three-case replay on native host runtimes, capturing complex library-mediated flows against inferred per-tool scopes. Similarly, where mobile analyzers such as PathSentinel~\cite{android-fs} evaluate applications against pre-existing platform access-control policies, MCP-FSD infers least-privilege boundaries directly from unstandardized tool descriptions and deployment configurations.

Description-code agreement, file-to-output flow, and authorization answer different questions. A description can promise to stage local files without specifying which directories are allowed. A diff response establishes that file bytes reached the client, while the repository-root policy states the scope against which the access is tested. MCP-FSD records the source candidate, policy, and execution transcripts together, connecting the declared operation to the tested access and returned bytes.

\textit{Agent security, prompt injection, and isolation.} Related work benchmarks harmful tool use~\cite{toolemu,agentdojo,agentharm,asb,rjudge,agentfuzz}, studies agent backdoors~\cite{badagent,watch-out-agents}, and attacks or defends model instruction-following~\cite{struq,jatmo,attention-defense,masterkey}. It also applies models to offensive-security tasks~\cite{pentestgpt,web-enabled-llms} and documents path traversal in the agent harness itself~\cite{claude-code-pt1,claude-code-pt2}. MCP-FSD tests the effects of separately launched MCP server processes using existing OS containment and deception primitives~\cite{ebpf-syscall,isolategpt,cloak-honey-trap,sandbox-runtime,nono}. Paired malicious replay and a benign control test the inferred policy (\S\ref{sec:design:dynamic}).

\section{Conclusion}
\label{sec:conclusion}

Filesystem capability laundering exposes a fundamental gap between client-side tool approval and unconfined server filesystem execution. MCP-FSD bridges this gap by coupling static screening, inferred least-privilege boundaries, and dynamic three-case validation across 21{,}097 cataloged servers. Real-world vulnerabilities in foundational tools---such as repository creation and staging flaws in Anthropic's Git server---demonstrate how ordinary tool interactions can inadvertently leak sensitive host credentials. These findings demonstrate the necessity of ecosystem-scale registry screening and explicit, runtime-enforced per-tool filesystem contracts.

\bibliographystyle{splncs04}
\bibliography{references}

\end{document}